\documentstyle[twocolumn,aps,prl]{revtex}
\input boxedeps  
\HideDisplacementBoxes 
 \SetRokickiEPSFSpecial 
\begin{document}

\draft
\preprint{TIT/HEP-284/COSMO-53}

\title{Critical behaviour in gravitational collapse of radiation
fluid\thanks{gr-qc/9503007, TIT/HEP-284/COSMO-53} \\
--- A renormalization group (linear perturbation) analysis ---}
\author{Tatsuhiko Koike\thanks{JSPS fellow,
e-mail: bartok@th.phys.titech.ac.jp}}
\address{Department of Physics, Tokyo Institute of Technology,
Oh-Okayama, Meguro, Tokyo 152, Japan}
\author{Takashi Hara\thanks{e-mail: hara@aa.ap.titech.ac.jp}
{\em and} Satoshi Adachi\thanks{e-mail: adachi@aa.ap.titech.ac.jp}}
\address{Department of Applied Physics, Tokyo Institute of Technology,
Oh-Okayama, Meguro, Tokyo 152, Japan}

\maketitle

\begin{abstract}
A scenario is presented, based on renormalization group (linear
perturbation) ideas,  which can explain the self-similarity and scaling
observed in a numerical study of
gravitational collapse of radiation fluid.
In particular, it is shown that the critical exponent $\beta$ and the
largest Lyapunov exponent ${\rm Re\, } \kappa$ of
the perturbation is related by
$\beta= ({\rm Re\, } \kappa) ^{-1}$.
We find the relevant
perturbation mode numerically, and obtain a fairly accurate
value of the critical
exponent $\beta \simeq 0.3558019$, also in agreement with that
obtained in numerical simulation.
\end{abstract}

\narrowtext


\paragraph*{1. Introduction. }
Gravitational collapse with formation of black holes is one of the
main problems of classical general relativity.
Recently,  Choptuik \cite{Cho93} discovered
a ``critical behaviour'' of the gravitational collapse of massless scalar
field by a numerical study. His result can be summarized as follows:
Let the initial distribution of the scalar field be parameterized
smoothly by a parameter $p$,
such that the solutions with the initial data
$p>p^*$ contain a black hole
while those with $p<p^*$ do not.
For several  one-parameter families investigated,
near-critical solutions ($p \approx p^*$) satisfy the following:
(i) the critical solution (i.e. $p = p^*$) is {\em universal}
in the sense that it
approaches the identical spacetime for all families,
(ii) the critical solution  has a discrete self-similarity, and
(iii) for supercritical solutions ($p > p^*$) the black hole mass satisfies
$M_{BH}\propto (p-p^*)^\beta$
and the critical exponent $\beta$, which is about 0.37, is universal for
all families. Abrahams and Evans \cite{AbEv93} found similar
phenomena in axisymmetric collapse of gravitational wave.

Evans and Coleman
\cite{EvCo94}  found similar phenomena with $\beta\simeq0.36$
in spherically symmetric collapse of radiation fluid,
in which case the self-similarity is not discrete but {\em
continuous}.
Employing a self-similar ansatz they also found a numerical solution
which fits the inner region of the near-critical solutions very well,
and suggested that linear analysis around the solution will be useful.

In this letter,
we present a scenario that explains the
critical behaviour observed in the
radiation fluid collapse. We directly
relate eigenvalues (Lyapunov exponents) of perturbations
of the self-similar solution to the critical exponent $\beta$,
using an argument of renormalization group transformation. The formulation
is general, and could be applied to other models with approximate
self-similarity.  We find the
eigenvalues of the perturbation by numerical analysis,
and find that the value of the exponent
$\beta$ predicted from our analysis matches very well with that observed
in \cite{EvCo94}.

We present our picture in sec. 2. After reviewing the equations of motion
in sec. 3, and the
self-similar solution in sec. 4, we confirm our picture to an
extent in sec. 5 by numerical study. Sec. 6 is for conclusions and
discussions.


\paragraph*{2. A Scenario based on renormalization group ideas.}
\label{RG}
We give a formalism for linear perturbations around a
self-similar spacetime from the point of view of renormalization
group, which proved to be extremely  successful in the study of
critical phenomena \cite{WK74}.
We have benefitted from the formulation of \cite{BK95}.
The argument is general but the
notation is so chosen as to pave the shortest path
to our analysis on radiation fluid.  We introduce the ``scaling
variable'' $s$ and the ``spatial coordinate'' $x$, which are related to
the time $t$ and the radial coordinate $r$ by $s \equiv - \ln (-t)$,
$x \equiv \ln (- r/t)$.
\paragraph*{2a. Renormalization Group and Linear Perturbations.}

Let $h=(h_1, h_2, ..., h_m)$ be functions of $s$ and $x$
which satisfy a partial differential equation
\begin{equation}
  \label{eq:PDE}
  L(h,\frac{\partial h}{\partial s},\frac{\partial h}{\partial x}) =0.
\end{equation}
Suppose that the PDE is invariant under the ``scaling transformation''
(translation in $s$) with $s_0 \in {\bf R}$
\begin{equation}
        h(s,x) \mapsto
        h(s+ s_0,  x).
\end{equation}
A renormalization group transformation (RGT)
$\hat R_{s_0}$ is a transformation on the space of functions of $x$,
\begin{equation}
        \hat{\cal R}_{s_0}: H  \mapsto H^{(s_0)} ,
\end{equation}
where
\begin{eqnarray}
        H(x) & \equiv & h(0,x), \\
        H^{(s_0)}(x) & = &
        h(s_0,  x).
\end{eqnarray}
Namely, one obtains $H^{(s_0)}$ by evolving the initial data $H$
at $s=0$ by the PDE to $s=s_0$.
$\hat {\cal R}_{s_0}$ forms a semigroup with parameter $s_0$, and we
denote its generator by $\widehat{D{\cal R}}$, i.e.
$ \displaystyle \widehat{ D{\cal R} }
        = \lim_{{s_0}\rightarrow 0}{(\hat{\cal R}_{s_0}-{\bf 1}) / s_0}$.
In this context, a  self-similar solution $h_{\rm ss}(s,x) = H_{\rm ss}(x)$
can be
considered as a {\em fixed point} of $\hat {\cal R}_{s_0}$ for any
$s_0$, and is
characterized
by  $\hat {\cal R}_{s_0} H_{\rm ss} = H_{\rm ss}$ or
$\widehat{D{\cal R}}H_{\rm ss}=0$.

The {\em tangent map} of $\hat{\cal R}_s$ at a fixed point $H_{\rm ss}$
is defined as a transformation on functions of $x$:
\begin{equation}
        \hat{T}_{s_0} F \equiv \lim_{\epsilon \rightarrow 0}
        \frac{ \hat{\cal R}_{s_0}
        (H_{\rm ss} + \epsilon F ) - H_{\rm ss} } {\epsilon} .
        \label{ind.tan}
\end{equation}
An {\em eigenmode} $F(x)$ of
$\displaystyle \widehat{DT} = \lim_{s_0 \rightarrow 0}
(\hat T_{s_0} - 1) / s_0$ is a function which satisfies
($\kappa \in {\bf C}$)
\begin{equation}
        \widehat{DT} F = \kappa F.
        \label{eq:pert0}
\end{equation}
These modes determine the flow of the RGT near the fixed point.
A mode with ${\rm Re \,} \kappa>0$, a {\em relevant mode},
is a flow diverging from
$H_{\rm ss}$, and one with ${\rm Re \,} \kappa<0$, an {\em irrelevant mode},
is a flow
converging to it. A ${\rm Re \,} \kappa=0$ mode is called {\em marginal}.


\paragraph*{2b. The critical solution. }

In this and the next subsections, we present a scenario which
explains the observed critical behaviour of radiation fluid, assuming
that there is a {\em unique} relevant mode with eigenvalue $\kappa$
(and, for simplicity, no marginal mode)
around the fixed point $H_{\rm ss}$.  This assumption is
confirmed to some extent in sections 3--5.

\begin{figure}
\begin{center}
{\BoxedEPSF{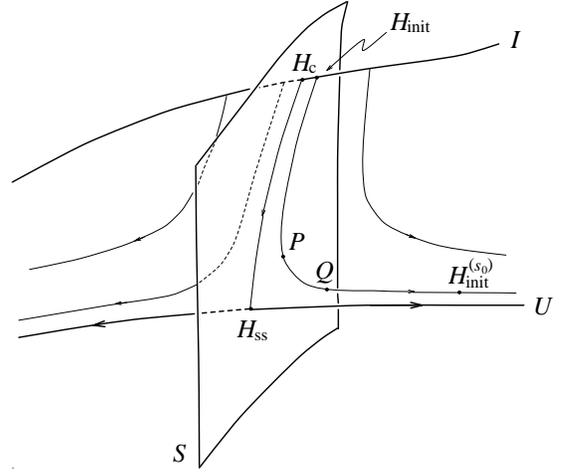 scaled 300} }
\end{center}
\caption{Schematic view of RG flows near $H_{\rm ss}$. }
\label{fig0}
\end{figure}

The assumption implies that the RG flow around the fixed point
is shrinking, except for the direction of the relevant mode (Fig.~\ref{fig0}).
There will be a ``critical surface'' or a ``stable manifold'' $S$ of the
fixed point $H_{\rm ss}$, of codimension one,
whose points will all be driven towards $H_{\rm ss}$.
There will be an ``unstable manifold'' $U$ of dimension one, whose
points are all driven away from $H_{\rm ss}$.
A one parameter family of initial data $I$ will in general intersect
with the critical surface,  and the intersection $H_{\rm c}$
will be driven to $H_{\rm ss}$ under the RGT:
\begin{equation}
          \lim_{s\rightarrow\infty}
      | H_{\rm c}^{(s)}(x) -  H_{\rm ss}(x) | =0 .
        \label{eq:crit}
\end{equation}
So, $H_{\rm c}$ is the initial data with critical parameter $p^*$.
The existence of a critical solution for an arbitrarily chosen
family of initial data thus supports the assumption of a unique
relevant mode \cite{footnote0}.

%


\paragraph*{2c. The critical behaviour. }
We now consider the fate of an initial data $H_{\rm init}$ in
the one-parameter family, which is  close to $H_{\rm c}$
 ($\epsilon = p - p^*$):
\begin{equation}
        H_{\rm init}(x) = H_{\rm c}(x) + \epsilon F(x) .
\end{equation}

We evolve this data to $s = s_0$ (chosen later):
it will first be driven
towards $H_{\rm ss}$ along the critical surface, but eventually be
driven away along the unstable manifold.
Using linear perturbations we have \cite{footnote1}:
\begin{eqnarray}
        H^{(s_0)}_{\rm init} & = &  \hat{\cal R}_{s_0} H_{\rm init}
        = \hat{\cal R}_{s_0} ( H_{\rm c} + \epsilon F )
        \nonumber \\
        & \simeq & H_{\rm c}^{(s_0)} + \epsilon \hat{T}_{s_0} F
        + O(\epsilon^2).
        \label{eq:lin1}
\end{eqnarray}
In the second term, only the relevant mode survives:
\begin{equation}
        \hat T_{s_0} F =
        \exp(s_0 \widehat{D T})F \simeq
        e^{\kappa s_0}  F_{\rm rel},
\end{equation}
where  $F_{\rm rel}$ is the component of the relevant mode
in $F$.
Due to (\ref{eq:crit}), we finally have (for large $s_0$ and
$x \, \raisebox{+0.1ex}{$<$} \hspace{-0.8em} \raisebox{-1ex}{$\sim$} \,  s_0$)
\begin{equation}
        H^{(s_0)}_{\rm init}(x) \simeq
        H_{\rm ss}(x) + \epsilon e^{\kappa s_0 }F_{\rm rel}(x).
        \label{eq:RG.result}
\end{equation}

Now we choose $s_0$ so that
the first and second terms in (\ref{eq:RG.result}) become
comparable, i.e.
\begin{equation}
        \label{eq:s_0value}
        \epsilon e^{ ({\rm Re \, } \kappa) s_0} = O(1).
\end{equation}
Now that the second term is of $O(1)$,
the data $H^{(s_0)}_{\rm init}$ differs from $H_{\rm ss}$
so much that
one can tell the fate of this data depending on the sign of $\epsilon$;
and if a
black hole is formed, the radius of its apparent horizon,
and thus its mass,  will be $O(1)$ measured in $x$.

Finally, we translate the above result back into our original coordinate
$(t,r)$.
The relation $r = e^{x-s}$ implies
that the radius of the apparent horizon, which is $O(1)$ measured in $x$,
is in fact $O(e^{-s_0})$ measured in $r$.
So we have from (\ref{eq:s_0value})
\begin{equation}
         M_{BH} \simeq O(e^{-s_0}) \simeq
	O(\epsilon^{ 1/ ({\rm Re \,}\kappa)}   ).
         \label{eq:MBH}
\end{equation}
Therefore the critical exponent is given exactly by
\begin{equation}
         \beta =  \frac{1}{{\rm Re \,} \kappa}  .
         \label{eq:beta.exp}
\end{equation}



\paragraph*{3. Equations of Motion. }
\label{SS}
The line element of any spherically symmetric spacetime is written as
\begin{equation}
\label{eq:LE}
        ds^2=-\alpha^2(t,r)dt^2+a^2(t,r)dr^2+r^2 (d\theta^2+\sin^2\theta
d\phi^2).
\end{equation}
The only coordinate transformation which preserves the form of eq.
(\ref{eq:LE}) is
\begin{equation}
\label{eq:CT}
        t\mapsto F^{-1}(t).
\end{equation}

We assume the matter content is perfect fluid having
energy-momentum tensor
$
        T_{ab}=\rho u_au_b+p(g_{ab}+u_au_b),
$
where $u^a$ is a unit timelike vector and $p=(\gamma-1)\rho$.
We only consider the radiation fluid $\gamma=4/3$ in this letter.
The components of $u^a$ can be written as
$u_t=a (1-V^2)^{-1/2}$ and $u_r=aV (1-V^2)^{-1/2}$,
where $V$ is the 3-velocity of fluid particles.
In terms of variables $s \equiv -\ln(-t), x \equiv \ln (- r/t)$,
and introducing
$        N \equiv \alpha a^{-1} e^{-x},
        A \equiv a^2,
        \omega \equiv 4 \pi r^2 a^2 \rho ,
$
we can write the
equations of the system in an autonomous form, which makes the scale
invariance of the system transparent:
\begin{eqnarray}
        & & \frac{A_{,x}}{A} = 1 - A + \frac{2 \omega}{1-V^2}
                \left (1 + \frac{V^2}{3} \right ) , \nonumber \\
        & & \frac{N_{,x}}{N}
                = -2 + A - \frac{2 \omega}{3}, \nonumber \\
        & & \frac{\omega_{,s} + (1 + N V) \omega_{,x}}
                        {\omega} +
                        \frac{4 \{ V V_{,s} + (N + V) V_{,x} \} }
                        {3 ( 1 - V^2) }
                        \nonumber \\
                & & \quad \; \;
                 - \frac{N V A_{,x}}{3A}
                + \frac{4 V N_{,x}}{3} + 2 N
                  \left ( 1 + \frac{4 \omega}{9 ( 1- V^2) } \right )
                        = 0 , \nonumber \\
        & & \frac{4 V \omega_{,s}
               + (4 V + N + 3 N V^2) \omega_{,x}}
                                 {\omega} \label{eq:EOM}
\\
        & & \quad \; \; +
             \frac{4 \{(1 + V^2) V_{,s} + (1 + V^2 + 2 N V) V_{,x} }
                                 {1 - V^2}
                                 \nonumber \\
        & & \; \; + \frac{N (1 - V^2) A_{,x}}{A}
                + 4(1 + V^2) N_{,x}
                + 2 N ( 1 + 3 V^2) = 0. \nonumber
\end{eqnarray}


\paragraph*{4.  The critical solution.}
One obtains self-similar spacetimes by assuming that $N$ and $A$
depend only on $x$: $N=N_{\rm ss}(x), A=A_{\rm ss}(x)$.
Conversely, it can be shown \cite{AHK95b} that one can
express any spherically symmetric self-similar spacetimes in that
form if one retains the freedom of coordinate transformation
(\ref{eq:CT}).
Then it follows from eqs. (\ref{eq:EOM}) that $\omega_{\rm ss}$
and $V_{\rm ss}$ are also
functions of $x$ only: $\omega=\omega_{\rm ss}(x), V=V_{\rm ss}(x)$.
We fix the coordinate system by
requiring that the sonic point (see below) be at $x=0$.

We require (i) that the
self-similar solution be analytic for all $x \in {\bf R}$, and (ii) as a
boundary condition that the spacetime and the matter are regular,
$A=1$ and $V=0$, at the center ($x = - \infty$).
As has been extensively studied by Ori and  Piran \cite{OP89},
the analyticity condition (i) requires that
the solution be analytic in particular around
the sonic point, where the velocity of the fluid particle seen
from the observer on the
constant $x$ line is equal to the speed of sound $1/\sqrt{3}$.
The sonic point is a singular point for the ODE's
satisfied by
self-similar solutions, and considering the power series
expansion, one can see that
the solutions are specified by one parameter,
say, the value of $V_{\rm ss}(0)$.
This, together with the regularity condition at the center (ii),
restricts $V_{\rm ss}(0)$ to have only discrete values.
We employ the Evans--Coleman
self-similar solution as our $H_{\rm ss}$ in the following.


\paragraph*{5. Perturbation.}
\label{pert}
Perturbation equations (\ref{eq:pert0}) are obtained by taking the first order
variation in eqs. (\ref{eq:EOM}) from the
Evans--Coleman solution $H_{\rm ss}$:
\begin{equation}
        h(s,x) = H_{\rm ss}(x) + \epsilon  h_{\rm var}(s,x),
        \label{eq:pert.exp1}
\end{equation}
where $h$ represents each of $(N, A, \omega, V)$.
We require $N_{\rm var}(s,0)=0$ to fix the coordinate freedom (\ref{eq:CT}).

We consider
eigenmodes of the form $h_{\rm var}(s,x) = h_{\rm p}(x)e^{\kappa s}$,
with $\kappa \in {\bf C}$ being a constant.  Substituting this form into
(\ref{eq:pert.exp1}), and then into (\ref{eq:EOM})
yields linear, homogeneous first order
ODE's for $(N_{\rm p}, A_{\rm p}, \omega_{\rm p}, V_{\rm p})$.

As in the case of self-similar solutions, we require
(i) that the perturbations are analytic for all $x \in {\bf R}$, and
(ii) that the perturbed spacetimes are regular at the center
($A_{\rm p}, V_{\rm p}$ are finite at $x = - \infty$).
The sonic point becomes  a regular singular
point for the perturbations.
It is not hard to see that
apart from the overall multiplicative factor,
the perturbation solutions which satisfy the analyticity condition (i)
are specified by one free parameter $\kappa$.
This, together with the regularity condition (ii) at the center, allows
only discrete values for $\kappa$ in general.

\begin{figure}
\begin{center}
{\BoxedEPSF{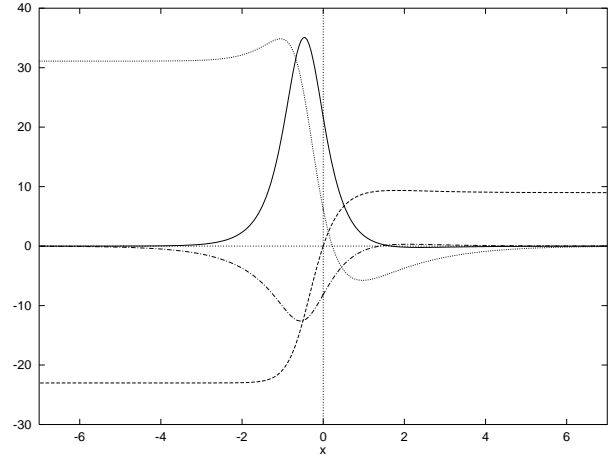 scaled 340} }
\end{center}
\caption{Profile of the eigenmode with the largest eigenvalue.
        Curves represent $A_{\rm p}$ (solid
        line), $N_{\rm ss}{}^{-1}N_{\rm p}$ (dashed),
        $\omega_{\rm ss}{}^{-1}\omega_{\rm p}$ (dotted),
        and $V_{\rm p}$ (dot-dashed).}
\label{fig2}
\end{figure}

Fig.~\ref{fig2} shows the profile of the largest
relevant eigenmode obtained numerically.
It has the eigenvalue $\kappa \simeq 2.81055255$; which corresponds to
the exponent value $\beta \simeq 0.35580192$, according to our scenario of
section~2.  This is in good agreement with the value of \cite{EvCo94}.

To further confirm our scenario,
we have checked that there are no  other relevant (or marginal)
eigenmodes in the range
$0 \leq {\rm Re \,} \kappa \leq 15, | {\rm Im \,} \kappa | \leq 14$.
We remark that there is an unphysical ``gauge mode''
at $\kappa \simeq 0.35699$ (in our gauge), which emerges from
a coordinate transformation
applied to the self-similar solution.
Due to the complicated structure of the equations of
motion, we have not found a beautiful argument which can restrict
possible eigenvalues (like that of \cite{Ch75} and
references therein).

\paragraph*{6. Conclusions and discussions.}
\label{conc}

In this letter we have presented a scenario based on the
renormalization group (linear perturbation) ideas, by which
the critical behaviour in the black hole formation in
the radiation fluid collapse is well understood.  In particular,
we have shown that the critical exponent $\beta$ is equal to
the inverse of the largest Lyapunov exponent ${\rm Re \,} \kappa$.
We have
performed a partial confirmation of the picture, and modulo
some assumptions about distributions of eigenmodes, have found
an accurate value  $\beta \simeq 0.35580192$, which is also close to
that reported in \cite{EvCo94}.

Complete confirmation of the scenario requires further study
of {\em local} and {\em global} structure of RG flows around the fixed
point $H_{\rm ss}$.  To establish the local picture, i.e. the
contraction
property of the RGT on the cospace of our relevant mode in {\em some}
neighbourhood of $H_{\rm ss}$, one could try to prove that the
eigenmodes form
a complete set of solutions and that {\em all} modes except our
relevant one are in fact irrelevant.
Establishing the global picture, which corresponds to
proving that the global critical surface exists and
the RG flow around it is as depicted in Fig. 1,
will pose another challenging problem
which will be beyond the scope of linear perturbations.
Complete knowledge of the RG flow could expand the  horizon of our
understanding the gravitational collapse.

It is expected that the critical behaviour in scalar field collapse,
where the self-similarity is discrete, can be understood in a
 manner similar to the analysis in this letter.
One can consider the critical solution as
a fixed point of the RGT $\hat {\cal R}_{s_0}$
with a suitably chosen $s_0$, and can perform
linear perturbation analysis  around it.
A work in this direction is now in progress.

Our intuition on gravitational collapse still seems to be heavily
based on few exact solutions, especially the limiting case of
pressureless matter.
The critical behaviour may
provide a different limiting case
that the final mass is small compared to the initial mass
for more realistic and wider range of matter contents.
It will be of great help to settle the problems in gravitational
collapse such as cosmic no hair conjecture.

\paragraph*{Acknowledgments.}
T. K. thanks Yasufumi Kojima for discussions.
We thank Akio Hosoya for reading our manuscript.
The research is supported in part by Japan Society for the Promotion of
Science (T. K.).
All the computations were performed on
several work stations
of Tokyo Institute of Technology.


\begin{thebibliography}{1}
\bibitem{footnote0}
If there are more than one relevant modes, the stable manifold of
$H_{\rm ss}$ will have codimension more than one; a generic
family of initial data will not intersect the stable manifold, thus
no sharp critical behaviour will be seen.
\end{thebibliography}
\end{document}